\newcommand{\DIMPY} {(C$_7$H$_{10}$N)$_2$CuBr$_4$\xspace}

\newcommand{\BPCB} {(C$_5$H$_{12}$N)$_2$CuBr$_4$\xspace}

\newcommand{\KCUC} {K$_2$CuSO$_4$Cl$_2$ \xspace}

\newcommand{\ii}{\mathrm{i}}
\newcommand{\dd}{\mathrm{d}}

\newcommand{\be}{\begin{equation} }
\newcommand{\ee}{\end{equation} }
\newcommand{\bea}{\begin{eqnarray} }
\newcommand{\eea}{\end{eqnarray} }

\newcommand{\NTENP}{Ni(C$_9$H$_24$N$_4$)(NO$_2$)(ClO$_4$)\xspace}

\newcommand{\CuDCl}{2(1,4-Dioxane)$\cdot$2(H$_2$O)$\cdot$CuCl$_2$\xspace}

\newcommand{\mb}[1]{\mathbf{#1}}

\documentclass[aps,prr,reprint,showpacs, citeonline]{revtex4-1}
\usepackage{amsmath,amssymb,bm}
\usepackage{graphicx}
\usepackage{xspace}
\usepackage{latexsym}
\usepackage{xcolor}
\usepackage{hyperref}
\usepackage{placeins}

\begin{document}
\title{Quantum critical dynamics and scaling in one-dimensional antiferromagnets} %
	
\author{Andrey Zheludev} 
\email{zhelud@ethz.ch} 
\affiliation{Laboratory for Solid State Physics, ETH Z\"{u}rich, 8093 Z\"{u}rich, Switzerland}

	\begin{abstract}
		For a number of quantum critical points in one dimension quantum field theory has provided {\em exact} results for the scaling of spatial and temporal correlation functions. Experimental realizations of these models can be found in certain quasi one dimensional antiferromagnetc materials. Measuring the predicted scaling laws experimentally presents formidable technical challenges. In many cases it only became possible recently, thanks to qualitative progress in the development of inelastic  neutron scattering techniques and to the discovery of new model compounds. Here we review some of the recent experimental studies of this type.		
	\end{abstract}
	
	\date{\today}
	\maketitle
\section{Introduction}

\subsection*{Antiferromagnets as models of quantum criticality}

Magnetic insulators have served as models of choice for studying criticality and scaling phenomena ever since the early days (the 1960s) \cite{Stanleybook,Collinsbook}. Deservedly so. These systems feature local degrees of freedom and very short-range interactions. This ensures that their behavior falls into one of the less trivial universality classes, rather than being of a Mean Field (MF) type. Furthermore, a number of very precise experimental techniques, such as  NMR and neutron scattering directly couple to the magnetic order parameter and can be conveniently employed to study spatial or temporal critical correlations. The extraordinary diversity of magnetic materials offers a variety of order parameters with Ising, Heisenberg or XY symmetries. Finally, real magnetic insulator compounds are often accurately described by very simple spin Hamiltonians with just a few experimentally measurable parameters. This allows for quantitative comparisons with theory and numerical simulations.

When later the focus shifted to {\em quantum} phase transitions (QPTs) and criticality \cite{Sachdevbook}, magnetic insulators started to play an even more important role. QPTs are driven by quantum, rather than thermal fluctuations \footnote{The Introduction section contains textbook information on quantum criticality and scaling. While there are plenty of good books and review articles on the subject, the author's personal preference is for Refs.~\cite{Sachdevbook,Continentinobook,Cadrybook}}. They occur at $T\rightarrow0$ upon a change of some extrernally controlled Hamiltonian parameter. An exceptionally convenient parameter for magnetic systems is an applied magnetic field.  As opposed to  chemical doping, it is ``clean'' and doesn't introduce any disorder. As opposed to pressure, it can be easily produced and finely controlled in the laboratory. Its coupling to the relevant degrees of freedom (spins) is direct and precisely known. Note that for experimental studies of field-driven quantum criticality the target material must be necessarily {\em antiferromagnetic} (AF). Indeed, for ferromagnets the applied field couples to the order parameter and takes the system further away from the critical point. 

 At a conventional thermodynamic phase transition dynamics and thermodynamics are completely disconnected. In contrast, at quantum criticality the dynamical exponent $z$, which relates the typical frequency of critical fluctuations to their size, $\omega \propto r^{-z}$, explicitly enters also the scaling laws for thermodynamic quantities. This calls for combined thermodynamic and spectroscopic studies. Here AF materials offer additional advantages. In these systems the critical vector (often the magnetic propagation vector in the ordered phase) is far removed in reciprocal space from structural Bragg peaks  and accoustic phonons. This allows uninhibited   momentum-resolved measurements of critical fluctuations and critical dynamics using neutron diffraction and spectroscopy \cite{Collinsbook}. A related issue is that the effective dimensionality of quantum critical points (QCPs) is augmented by the dynamical exponent: $d_\mathrm{eff}=d+z$ \cite{Sachdevbook}. The latter is always positive, so the system is pushed up towards the critical dimension and beyond. To avoid the rather uninteresting MF scaling regime one has to pick materials with a low physical dimension $d$ to start with. Fortunately, antiferromagnetic spin chain-, ladder- and layered compounds abound and make for great prototype systems. It is the studies of quantum critical dynamics in such (quasi) low dimensional AFs that this review is primarily focused on. By no means is this a new subject. Most of the theory was completed in the 1990s. Quite a few groundbreaking experiments were done shortly thereafter. Nevertheless, numerous key measurements were enabled only by very recent progress in materials and experimental techniques. Long-standing theoretical predictions finally got confirmation in the laboratory.

\subsection*{Theoretical expectations and experimental tasks}
Assuming a suitable prototype system is found, let's briefly summarize what precisely we would like to see experimentally. We shall distinguish two scenarios. In the first, the quantum AF is tuned to the criticality by an external uniform magnetic field $H$, the QCP ocurring at $H=H_c$. In this category are field-induced saturation transitions in almost any AFs, soft mode transitions in gapped spin chains and ladders, and various realizations of the transverse-field Ising model. The ``distance'' from the QCP is given by the temperature $T$ and by $h=H-H_c$. Assuming that at the quantum critical point $d+z$ is below the upper critical dimension, the most commonly measured properties (specific heat, magnetization and uniform susceptibility) can be written in  the following scaling forms \cite{Continentinobook}:
\bea
C_V(T,h)&=&T^{d/z}\,\mathcal{C}(\tilde{h}),\\
M(T,h)&=&T^{\frac{(d+z)\nu-1}{z\nu}}\,\mathcal{M}(\tilde{h}),\\
\chi(T,h)&=&T^{\frac{(d+z)\nu-2}{z\nu}}\,{X}(\tilde{h}).\label{scalingthermo}
\eea
Here in the RHS of each equation is the corresponding scaling function of a single variable $\tilde{h}=hT^{-1/{z\nu}}$, which is the scaled field. If $z\nu=1$ as is often the case \footnote{An important and common case of $z\nu\equiv 1$ is when the Zeeman term commutes with the rest of the Hamiltonian  \cite{Sachdev1994-2}, as, for example, in the Heisenberg model.} it becomes convenient to define the scaled field as a dimensionless quantity  $\tilde{h}=g\mu_B h/k_B T$.   

The dynamical properties of the system are quantified by the magnetic dynamic structure factor $\mathcal{S}(\mathbf{q},\omega)$, which is the temporal and spatial Fourier tzransform of the spin pair correlation functions and can be directly measured in inelastic neuron scattering experiments \cite{Squiresbook,Loveseybook}. It is expected to scale as:
\be
S(\mathbf{q},\omega,T,h)=T^{-x}\,{\mathcal{S}}_T(\tilde{q},\tilde{\omega},\tilde{h}),\label{SQWT}
\ee
where $x=\frac{\gamma+z\nu}{z\nu}$, the wave vector $\mathbf{q}$ is measured relative to the critical wave vector $\mb{q}_c$, $\tilde{q}=q/T^{1/z}$ is the scaled wave vector and $\tilde{\omega}=\hbar \omega/k_BT$ is the scaled frequency. For practical purposes, one oftem measures the {\em local} dynamic structure factor obtained from $\mathcal{S}(\mathbf{q},\omega)$ by integrating over momentum transfer. It scales as:
\be
{S}(\omega,T,h)=T^{-x+d/z}\,{\mathcal{S}}_T(\tilde{\omega},\tilde{h}),
\ee
An alternative scaling form for the dynamic structure factor is useful for measuring the frequency dependence at a constant temperature, as is usually done in neutron scattering experiments:
\be
{S}(\mathbf{q},\omega,T,h)=\omega^{x}\,{\mathcal{S}}_\omega(q/\omega^{1/z},T/\omega,h/\omega^{1/{z\nu}}).
\ee
Dynamics can also be probed by NMR. The measured relaxation rate $t_1$ is inversely proportional to $\mathcal{S}(\omega)$ for $\omega \rightarrow 0$ and thus scales as:
\be
{t}_1(T,h)=T^{x-d}\,{\tau}(\tilde{h}),
\ee

The second scenario is when a quantum phase transition is driven not by the applied field, but by some other (often experimentally inaccessible) parameter and the system happens to ``automatically'' be at quantum criticality. Often it may remain critical in a wide range of applied magnetic fields. The best known example is the antiferromagnetic Heisenberg  $S=1/2$  chain, which has a quantum-critical ground state already in zero field but also in applied fields all the way up to saturation \cite{Tsvelikbook,Giamarchibook}. Very similar physics is realized in a Heisenberg $S=1$ (Haldane) spin chain or a $S=1/2$ spin ladder in a magnetic field beyond the soft mode transition. A rather different example would be a a bond-alternating $S=1$ chain where the dimerization is just strong enough to close the Haldane gap, but not enough to re-open a dimer-gap\cite{Affleck1987-2,Yamamoto1995}. An even more exotic case is that of a spin ladder that is tuned to the so-called $SU(2)_2$ WZNW quantum critical point by cyclic exchange interactions\cite{Lake2010}.  In all these instances the magnetic field is not what controls criticality. The scaling relations are as above but without any explicit field variable on either the LHS or RHS, albeit with possibly field-dependent scaling functions and exponents.

For either scenario, the first task  for the experimentalists is to verify that the thermodynamic quantities and the dynamic structure factor scale at all. The next step is to measure the power laws for a comparison with theoretical predictions. It is also important to verify that spectroscopy and calorimetry experiments yield consistent sets of scaling exponents. Often specific theoretical predictions exist also for the scaling functions, which may be completely or at least partially universal. Measuring these functions experimentally is then particularly gratifying, because a {\em quantitative} comparison with theory can be done {\em with almost no quantitative knowledge about the material under study}.

\subsection*{Technical challenges}

Despite all the advantages offered by quasi low dimensional AFs, experiments aimed at measuring their quantum critical properties  are anything but easy. The main issue is that for the transition to be driven by reasonable magnetic fields (up to, say, 10~T) the energy scale of magnetic exchange interactions needs to be similarly low, of the order of a few meV. In many ``traditional'' low-dimensional quantum AFs realized in 3d-metal oxides (byproducts of the frantic search for HTSC cuprates in the 1990s)  the magnetic energy scale is simply too high. This is one reason why much of the resent progress was done on 3d-metal-halide/ organic complexes, where exchange interactions  are typically smaller \cite{Yankova2011}.

Another reason why the choice of model compound is crucial is that the energy range of the relevant critical fluctuations is typically another order of magnitude or so smaller than the exchange constants, i.e., a few hundred $\mu eV$. Any unwanted symmetry breaking terms in the spin Hamiltonian of the system (residual 3-dimensional interactions, anisotropy, magneto-elastic coupling, etc.) have to be smaller still. The initially long list of potential prototype materials rapidly shrinks to leave only a handful of suitable species, many of them only recently discovered. 

The low energy scale of critical fluctuations also imposes severe constraints on spectroscopic experiments. If quantum critical dynamics is to be measured using inelastic neutron scattering, an energy resolution of as little as 10~$\mu eV$ may in certain cases be required. Not long ago such experiments would be unthinkable. It is only in recent years that instruments providing such a high resolution at a reasonable flux became available to the experimentalist. Below we shall review several of such recent neutron measurements.

\section{Tomonaga-Luttinger spin liquids}
In the first case study we look at {\em  gapless one-dimensional Heisenberg AFs with a linear spectrum}. For the simplest example, that of a $S=1/2$ spin chain, the Jordan-Wigher transformation exactly maps the spin Hamiltonian onto that of interacting fermions\cite{Tsvelikbook}. As first suggested by Haldane \cite{Haldane1980}, a fermionic description of the low-energy physics is actually valid in all cases \cite{Tsvelikbook, Giamarchibook}. It also holds in applied magnetic fields. Moreover, systems like spin ladders that may be gapped in zero field will fall into this category in fields exceeding the field $\mu_0H_{c}$ at which the gap closes. The fermionic correspondence ensures {\em universality} in the low-temperature properties and dynamics. Indeed, in one dimension, regardless of the details of their interaction, fermions with a linear dispersion form the so-called Tomonaga-Luttinger liquid\cite{Tsvelikbook}. This is a {\em quantum critical state} fully characterized by just two constants: the Fermi velocity $u$ and the so-called Luttinger parameter $K$\cite{Tsvelikbook,Giamarchibook}. The latter is defined by interactions: $K>1$ corresponds to attraction and $K<1$ to repulsion. 

Due to the fermionic description, gapless one-dimensional spin systems with a linear spectrum are referred to Tomonaga-Luttinger spin liquids (TLSL).  The corresponding velocity and the Luttinger parameter will depend on the details of the spin Hamiltonian and the applied magnetic field. The $S=1/2$ Heisenberg chain in zero field corresponds to $K=1/2$. The interactions remaining repulsive all the way to the saturation field $H_\mathrm{sat}$, whre $K$ finally reaches unity \cite{Giamarchibook}. In magnetized Heisenberg $S=1/2$ spin ladders \cite{Hikihara2001,Schmidiger2012}, as well as  general XXZ spin chains \cite{Giamarchibook}, there is also a possibility of attractive interactions ($K>1$). Since the TLSL is critical and the spectrum is linear, all the low-temperature thermodynamic properties show scaling behavior with $z=1$. The thermodynamics is similar to that of a metal, with linear specific heat and constant susceptibility:
\bea
C_V(T)&=& Na k_B\frac{ k_B T }{6 \hbar u}\label{TLSLCV}\\
\chi(T)&=& \mu_0\frac{Na}{V}\frac{(g\mu_B)^2}{2\pi} \frac{K}{\hbar u},\label{TLSLCHI}
\eea
per magnetic ion, $a$ being the chain lattice constant and $N$ the number of spins in volume $V$.
Of course, the similarity doesn't extend to dynamics, since the Landau quasiparticle picture breaks down in one dimension. Fortunately, for the TLSL  the exponent and scaling function for the dynamic structure factor\cite{Schulz1996,Starykh1998,Giamarchibook} are known exactly:
\bea
&{S}(q,\omega,T)\propto T^{\frac{1-4K}{2K}}\,\left(1+\frac{1}{\exp(\hbar\omega/k_BT)-1}\right)\times\nonumber\\
&\times \mathrm{Im}\left\{\frac{\Gamma\left(\frac{1}{8K}-\ii \hbar\frac{ u{q}/a+{\omega}}{4\pi k_BT}\right)\Gamma\left(\frac{1}{8K}-\ii \hbar\frac{ u\tilde{q}/a-{\omega}}{4\pi k_BT}\right)}{\Gamma\left(1-\frac{1}{8K}-\ii \hbar\frac{ u\tilde{q}/a+{\omega}}{4\pi k_BT}\right)\Gamma\left(1-\frac{1}{8K}-\ii \hbar\frac{ u{q}/a-{\omega}}{4\pi k_BT}\right)}\right\},\label{TLSL:Sqw}
\eea
which obeys to Eq.~\ref{SQWT} without explicit field variables.

\subsection{The $S=1/2$ Heisenberg chain}
\subsubsection{Classic results}
\begin{figure}[h!]
	\centering
	\includegraphics[width=\columnwidth]{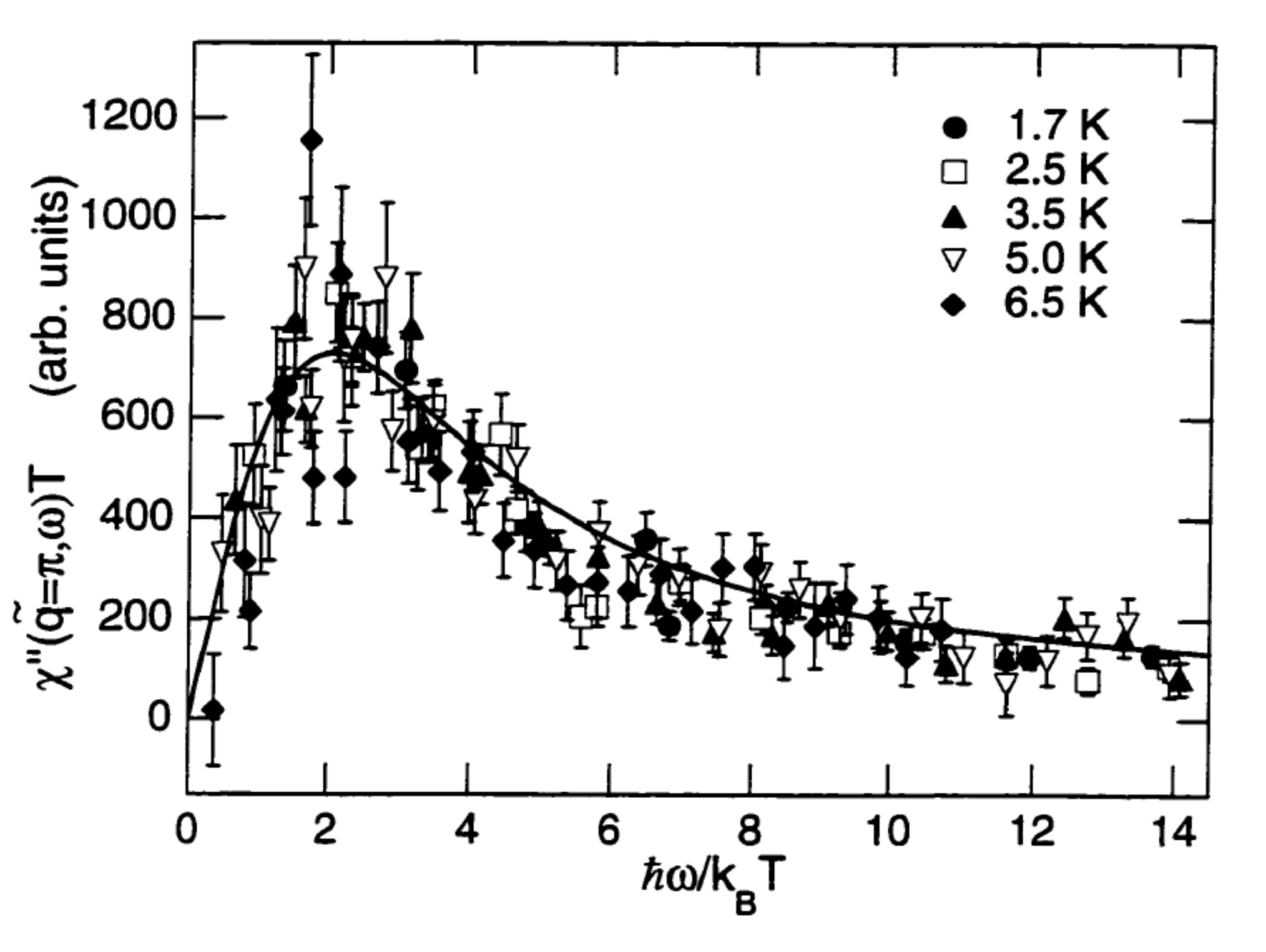}
	\caption{From Ref~\cite{Dender1997thesis}: Scaled critical dynamical structure factor measured in Cu-benzoate at several temperatures plotted against scaled frequency (symbols). The solid line is Eq.~\ref{TLSL:Sqw} with $K=1/2$. \label{Dender}}
\end{figure}

The first experimental tests of scaling in TLSLs  were done on $S=1/2$ Heisenberg spin chain compounds. Thermodynamic studies of prototypical compounds are now textbook material. Particularly useful were measurements on organic species such as Cu-benzoate\cite{Dender1997} and Cu-PZN \cite{Hammar1999}. The first systematic exploration of finite-$T$-scaling of  $\mathcal{S}(\mb{q},\omega)$ is particularly worth mentioning  in the context of the present review. These pioneering experiments on the organic spin-chain compound Cu-benzoate were carried out by D. Dender and C. L. Broholm over two decades ago \cite{Dender1997thesis}. Fig.~\ref{Dender} shows  the scaling plot for the imaginary part of susceptibility measured at the one-dimensional AF zone-center ${\mb{q}}=0$ at different temperatures. That quantity is directly related to the dynamic structure factor through the fluctuation-dissipation theorem: $\chi''(\mb{q},\omega)=\pi[1-\exp(-\hbar \omega/k_B T)]{S}(\mb{q},\omega)$. The solid line is a fit to Eq.~\ref{TLSL:Sqw} with $K=1/2$. The quality of the data collapse and agreement with theory are remarkable. 

More frequently cited is a later study  on another compound, namely KCuF$_3$ \cite{Lake2005}. This material is a poor approximation of a one-dimensional magnet. Inter-chain interactions are rather strong and  result in 3-dimensional magnetic long-range order already at $T_N=39$~K and achieves a sublattice magnetization of about 0.5~$\mu_B$ per Cu$^{2+}$ at low temperatures\cite{Hutchings1969}. Consequently, the low-temperature thermodynamics is nothing like that of an ideal spin chain. Nevertheless, one-dimensional scaling behavior given by Eq.~\ref{TLSL:Sqw} was shown to dominate the excitation spectrum at energy transfers exceeding about 25~meV$\sim 260$~K \cite{Lake2005} . This revelation highlights the key advantage of spectroscopy over static or low-frequency experiments.  If there are terms in the Hamiltonian that lead the system away from quantum criticality, one can still probe quantum critical dynamics and measure the predicted scaling relations at high energies. Thermodynamic  and NMR measurements are unable to employ a similar ``trick''.

\subsubsection{Diluted spin chains in zero field}
A rather recent development is the study of critical dynamics in  $S=1/2$ Heisenberg chain materials with spin-dilution. One replaces a small fraction of  the $S=1/2$ ions with integer-spin ones. Due to magnetic screening, the latter  have the same effect as $S=0$ defects  \cite{Eggert1992}. Since each defect breaks the continuity of the system entirely, the physics is that of TLSLs confined to finite-size ``boxes''. This problem has been studied extensively theoretically\cite{Eggert1992,Fujimoto2004,Sirker2008}. In a nutshell: due to $z=1$, the correlation length in a TLSL goes as $\hbar u/k_BT$. For a spin chain of length $L a$, the relevant parameter is $L a k_BT/\hbar u$. In fact, all thermodynamic properties can be written as universal functions of this ratio. This is the so-called $LT$ scaling. Its consequences are directly observable in experiments as long as one keeps in mind that in a real spin chain material with site dilution there is a statistical distribution of chain-fragment length. Most measurements of the resulting behavior  were carried out on linear-chain cuprates SrCuO$_2$ and SrCu$_2$O$_3$ diluted with integer-spin transition metal ions on the Cu site. Not only the uniform magnetic susceptibility \cite{Karmakar2015,Simutis2016,Simutis2017,Simutisthesis}, but also the staggered susceptibility, deduced from 3-dimensional ordering temperatures due to weak inter-chain coupling \cite{Simutis2016,Simutisthesis}, were found to be fully in agreement with $LT$ scaling predictions. 

\begin{figure}[h!]
	\centering
	\includegraphics[width=\columnwidth]{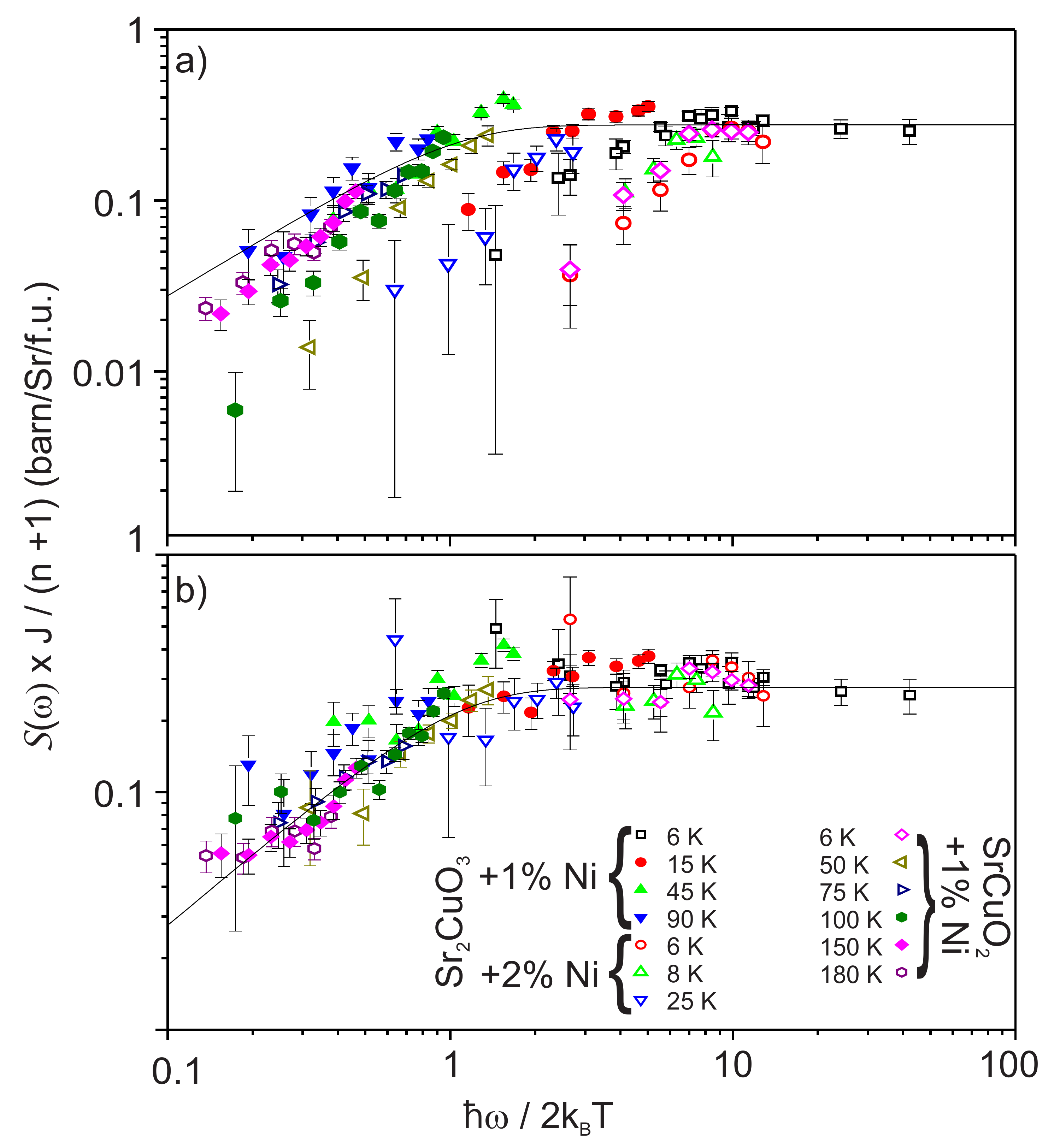}
	\caption{From Ref~\cite{Simutis2017}: Local dynamical structure factor measured in site-diluted spin chain compounds at different temperatures using time of flight (open symbols) and 3-axis (solid symbols) neutron spectroscopy techniques. (a) $q$-integrated neutron intensities as measured. (b) Same data scaled by the envelope function to factor out the spin pseudogap.  The solid line is the exact scaling function for an ideal $S = 1/2$ Heisenberg chain, plotted with no adjustable parameters. The data are from the SEQUOIA time-of-flight (TOF) instrument (Oak Ridge National Laboratory, USA), MAPS and MERLIN TOF-spectrometers (Rutherford Appleton Laboratory, UK), IN22 and IN8 3-axis spectrometers (Institut Laue-Langevin, Grenoble, France), the PUMA 3-axis spectrometer (Forschungsneutronenquelle Heinz Maier-Leibnitz, TU Muenchen, Germany) and the 4F2 3-axis machine at (Laboratoire Leon Brillouin, CEA-CNRS, Saclay, France) . \label{SCO}}
\end{figure}

The most elegant experimental result on diluted spin chains pertain to the ``hidden'' finite-$T$ scaling of the local dynamic structure factor \cite{Simutis2013,Simutis2017,Simutisthesis}. Any length-$L$ chain fragment will have a spin gap  $\Delta=\Delta_0/L$, where $\Delta_0\sim 3.65  J$, $J$ being the exchange constant \cite{Eggert1992}. Due to a distribution of fragment lengths in a macroscopic sample, the spectrum will have a pseudogap. Obviously the dynamic scaling relations for an infinite spin chain (Eq.~\ref{TLSL:Sqw}) will no longer hold. However, it can be argued that the local dynamic structure factor $\mathcal{S}(\omega)$  in this case can be written as a product of that for defect-free chains $\mathcal{S}_\infty(\omega)$ and an ``envelope'' function defined by the defect concentration $x$ \cite{Simutis2013,Simutisthesis}: 
\be
{S}(\omega,T)={S}_\infty(\omega,T)\times\, \left(\frac{x\Delta_0}{\hbar\omega}\right)^2\sinh^{-2}\left(\frac{x\Delta_0}{\hbar\omega}\right).
\ee
Factoring out this envelope function should restore the scaling behavior. Experimentally, this is indeed the case. Fig.~\ref{SCO}a~\cite{Simutis2017} shows  $q$-integrated inelastic neutron scattering intensity  measured {\em on the absolute scale} at several temperatures in SrCuO$_2$ and SrCu$_2$O$_3$  with different defect concentrations, plotted against $\tilde{\omega}$. Obviously there is no data collapse between different temperatures, so the scaling is broken by defects. However, normalizing these data by the calculated respective envelope functions puts them on a single curve (Fig.~\ref{SCO}b). Moreover, the normalized data agree well with the known scaling function for $K=1/2$: $\mathcal{S}_T(\tilde{\omega})\propto \tanh(\tilde{\omega})$, shown in a solid line. For a Heisenberg spin chain even the proportionality coefficient in the latter formula is known exactly. Thus the agreement with experiments in Fig.~\ref{SCO}b is obtained {\em without any adjustable parameters}.

\subsubsection{Applied fields}
The experiments reviewed above were all performed in zero applied fields. For this reason, the restrictions that using realistic magnetic fields imposes on materials choice and energy resolution of spectroscopic measurements do not apply. As soon as one attempts to study dynamic scaling in in external fields large enough to substantially modify the Luttinger parameter $K$ of the spin chain, these considerations become paramount. 

\begin{figure}[h]
	\unitlength1cm
	\includegraphics[width=\columnwidth]{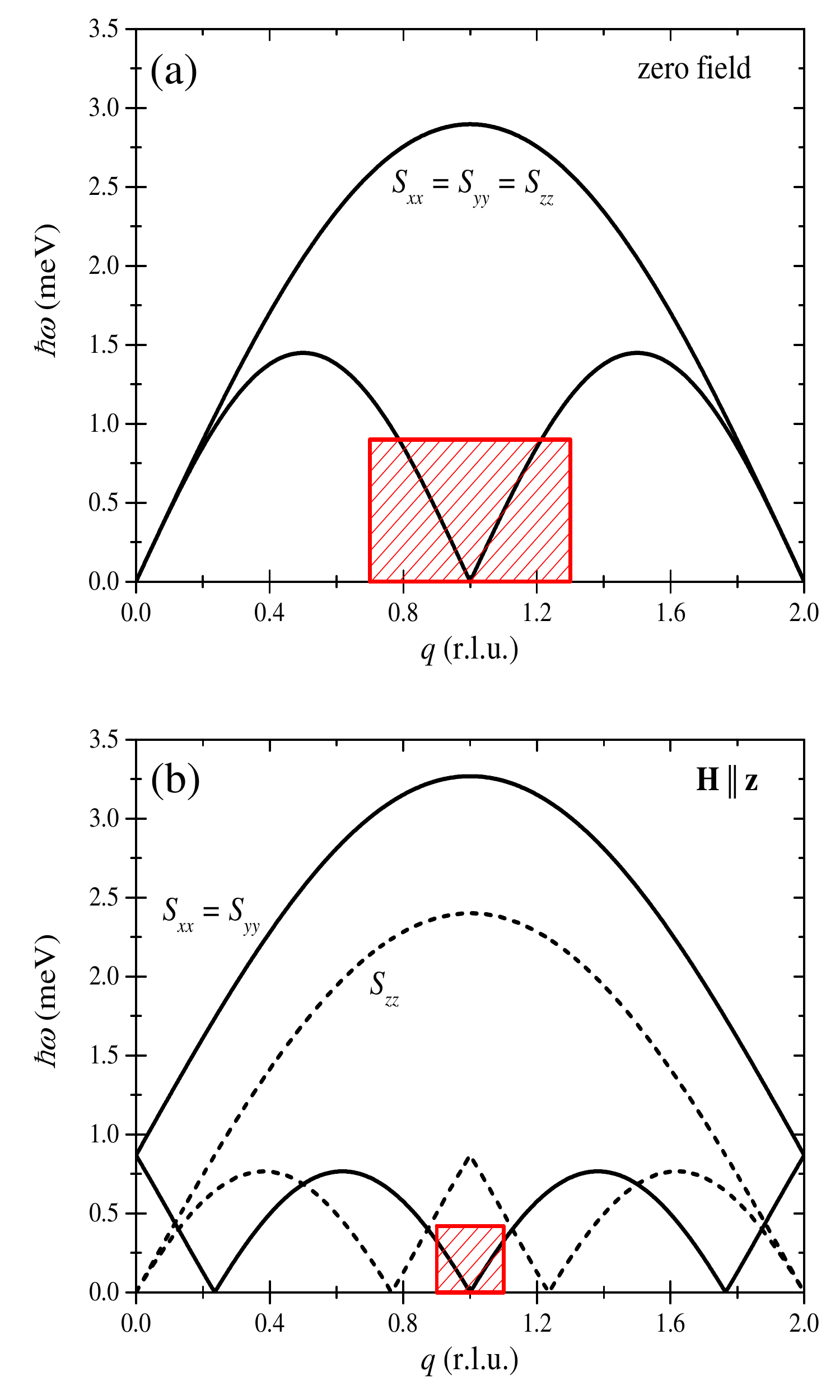}
	\caption{From Ref.~\cite{Haelg2015-2}: A sketch of the spin excitation spectrum of an AF Heisenberg $S=1/2$ chain in zero magnetic field (a) and in a magnetic field half way to saturation (b). The numerical values correspond to \CuDCl, where the saturation field is about 15~T. The approximate region where correlations are governed by TLSL physics is hatched. \label{Diox1}}
\end{figure}

Consider the schematic in Fig.~\ref{Diox1}a. In zero field the dynamic structure factor is isotropic and confined to within the bounds of the two-spinon continuum.
TLSL behavior is expected for a fairly wide energy and momentum range where the spinon dispersion is approximately linear (shaded rectangle). In applied fields the spectrum is modified.
The continua of spin fluctuations transverse and parallel to the field become distinct as shown in Fig.~\ref{Diox1}b. 
In the low-energy limit both polarization channels show TLSL scaling, but with different values of the scaling exponents and at different critical wave vectors (longitudinal excitations become incommensurate) \cite{Giamarchibook}. 
Using polarized neutrons to separate the two spectral components is impractical,
since that technique incurs a severe intensity penalty. To measure any one group of critical fluctuations one has to considerably shrink the measurement window to the immediate proximity of the corresponding critical wave vector (shaded area in Fig.~\ref{Diox1}b for transverse excitations).
A large window and an appreciable change of $K$ require high fields that substantially magnetize the spin chain. Yet a realistic field limit of a split-coil magnet used in a neutron experiment is typically 10~T depending on the setup.
To measure at 50\% saturation field, for instance, the exchange constant $J$ must be  of the order of 1~meV. This gives a measurement window of only about 0.5~meV and calls for a very high energy resolution of at least 50~$\mu$eV.

\begin{figure}[h]
	\includegraphics[width=\columnwidth]{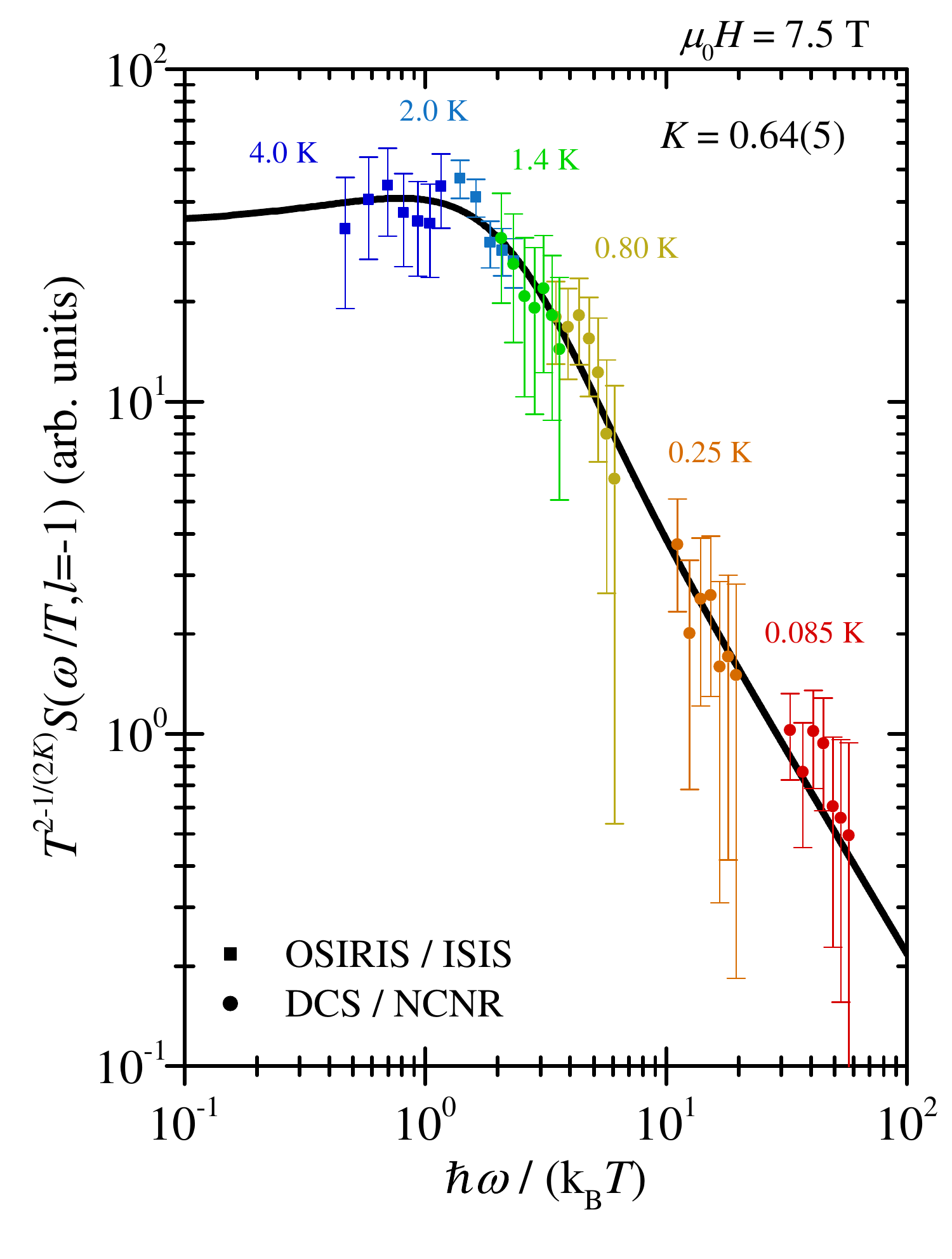}
	\caption{From Ref.~\cite{Haelg2015-2}: Scaling plot for $\mathcal{S}(0,\omega)$ measured in \CuDCl assuming $K=0.64$. The solid line is Eq.~\ref{TLSL:Sqw}. the data were taken at the OSIRIS backscattering spectrometer (Rutherford Appleton Laboratory, UK) and the DCS time-of-flight instrument (National Institute of Standards and Technology, Maryland, USA)  \label{Diox2}}
\end{figure}

The required resolution is just about achievable on modern time of flicht neutron instruments, particularly those that employ the backscattering principle. The corresoponding experiments~\cite{Haelg2015-2,Haelgthesis} were performed on the  spin chain compound \CuDCl for which very large single crystals can be grown. The material has $J=0.92$~meV and reaches saturation at $\mu_0H\sim 15$~T \cite{Hong2009}. Unfortunately, the spin chains have a staggered $g$-tensor which opens a small gap in applied fields. As with inter-chain interactions on KCuF$_3$, the effects of this unwanted small term in the Hamiltonian can be avoided by measuring at higher energy transfers. Although this further shrinks the already narrow measurement window, a reasonably good scaling plot can be obtained (Fig.~\ref{Diox2}, from Ref.~\cite{Haelg2015-2,Haelgthesis}). The $y$-axis scaling exponent $2-1/2K$ for this case is determined experimentally, to maximize the overlap between data collected at different temperatures. This is achieved at $K=0.64(5)$ as compared to the theoretical value $K=0.65$ for this value of applied field. The solid line is Eq.~\ref{TLSL:Sqw} with an arbitrary overall scale factor and shows good agreement with the measurement.

\subsection{Magnetized spin ladders}
Under all circumstances a Heisenberg $S=1/2$ chain corresponds to $K\le 1$, i.e., to repulsive Fermions. As mentioned above, a partially magnetized S=1/2 Heisenberg AF {\em spin ladder} can show $K>1$ TLSL behavior \cite{Hong2010,Schmidiger2012,Ninios2012,Jeong2013,Povarov2015,Jeong2016}. This is not always the case. The much-studied strong-rung spin ladder system \BPCB (BPCB) has $K<1$ in the entire range of applied fields between $H_c$ and $H_\mathrm{sat}$ \cite{Bouillot2011,Jeong2016}. In contrast, it's strong-leg sister compound \DIMPY (DIMPY) is an attractive TLSL in any magnetic fields beyong the gap-closure transition. This is an exceptionally well-characterized material with $J_\mathrm{leg}=1.42$~meV,  $J_\mathrm{rung}=0.82$~meV \cite{Hong2010,Schmidiger2011,Schmidiger2012,Schmidigerthesis} and infinitesimal inter-ladder coupling $J'=6~\mu$eV \cite{Schmidiger2012}. Density Matrix Renormalization Group (DMRG) calculations based on these values brilliantly reproduce the measured field-evolution of thermodynamics\cite{Schmidiger2012,Jeong2016,Schmidigerthesis} and the full spin excitation spectrum \cite{Schmidiger2012,Schmidiger2013-2,Schmidiger2013,Schmidigerthesis}. The velocity extracted from specific heat measurements using Eq.~\ref{TLSLCV} and the Luttinger parameter determined from Eqs.~\ref{TLSLCV} and \ref{TLSLCHI} \cite{Schmidiger2012,Ninios2012,Jeong2016,Schmidigerthesis}, as well as from measurements of the NMR relaxation rate\cite{Jeong2013} and of 3-dimensional ordering \cite{Schmidiger2012,Schmidigerthesis}, which occurrs at very low temperatures due to the presence of $J'$, all confirm the DMRG result: in DIMPY $K=1$ at $\mu_0H_c=2.62$~T, steady increases reaching $K\sim 1.3$ at 15~T and then again decreases to $K=1$ at $\mu_0H_\mathrm{sat}=29$~T.  

\begin{figure}[h!t]
	\centering
	\includegraphics[width=1\columnwidth]{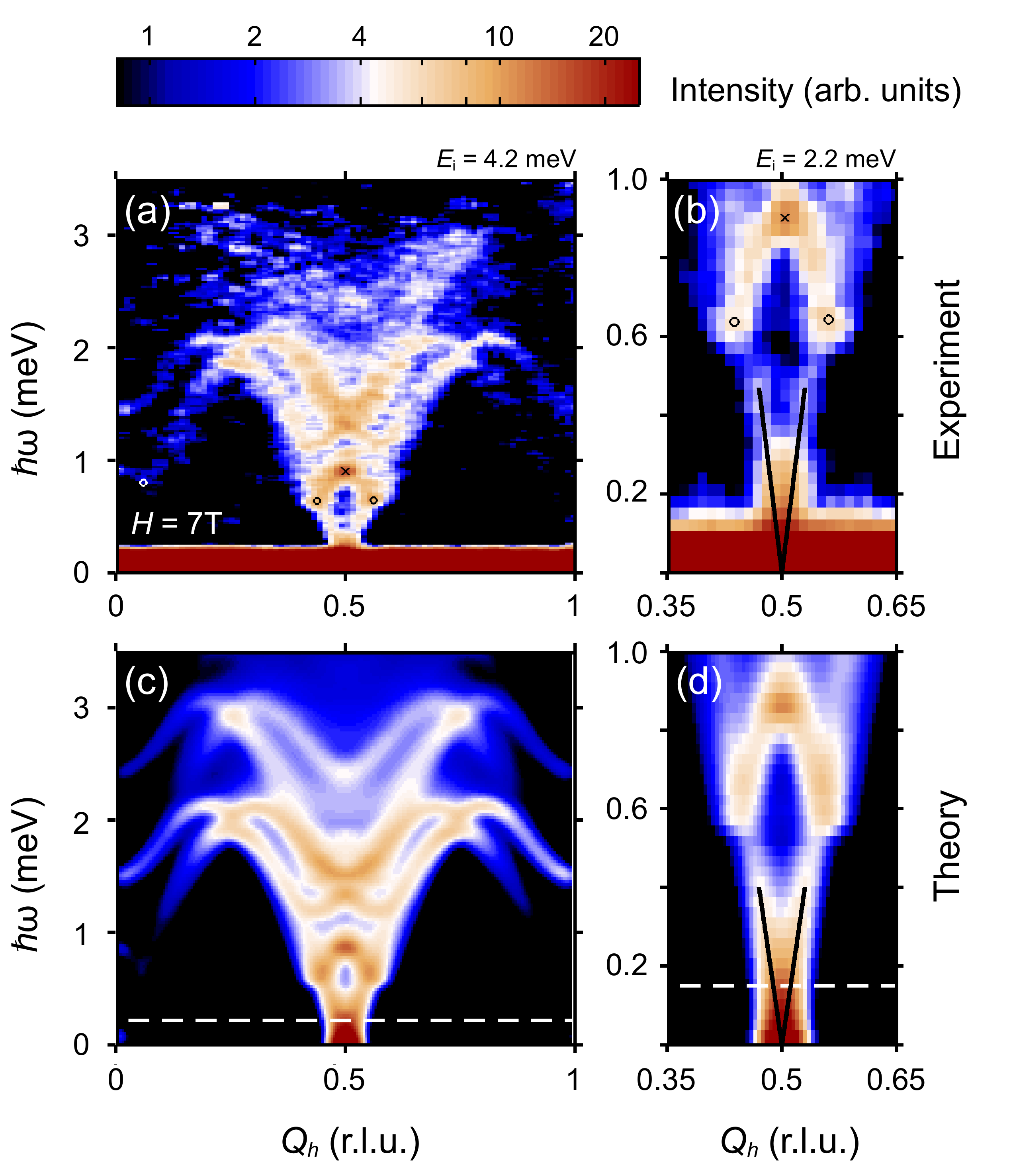}
	\caption{From Ref.~\cite{Schmidiger2013-2} Magnetic dynamic structure factor DIMPY in a magnetic field $\mu_0H = 7$~T. Inelastic neutron data were measured at $T=70$~mK in the low-resolution
		(a: $E_i = 4.2$~meV) and high resolution (b: $E_i = 2.2$~meV) modes.
		(c) and (d) are DMRG calculations, convoluted with experimental resolution.
		Dashed lines indicate the onset of the elastic incoherent background in the experiment.
	Lines indicate the dispersion cone of the gapless excitations. All critical spin fluctuations of the TLSL are confined to below 0.4~meV. The data are from the LET time-of-flight spectrometer (Rutherford Appleton Laboratory, UK).\label{DIMPY1}}  
\end{figure}

\begin{figure} \begin{center}
		\includegraphics[width=\columnwidth]{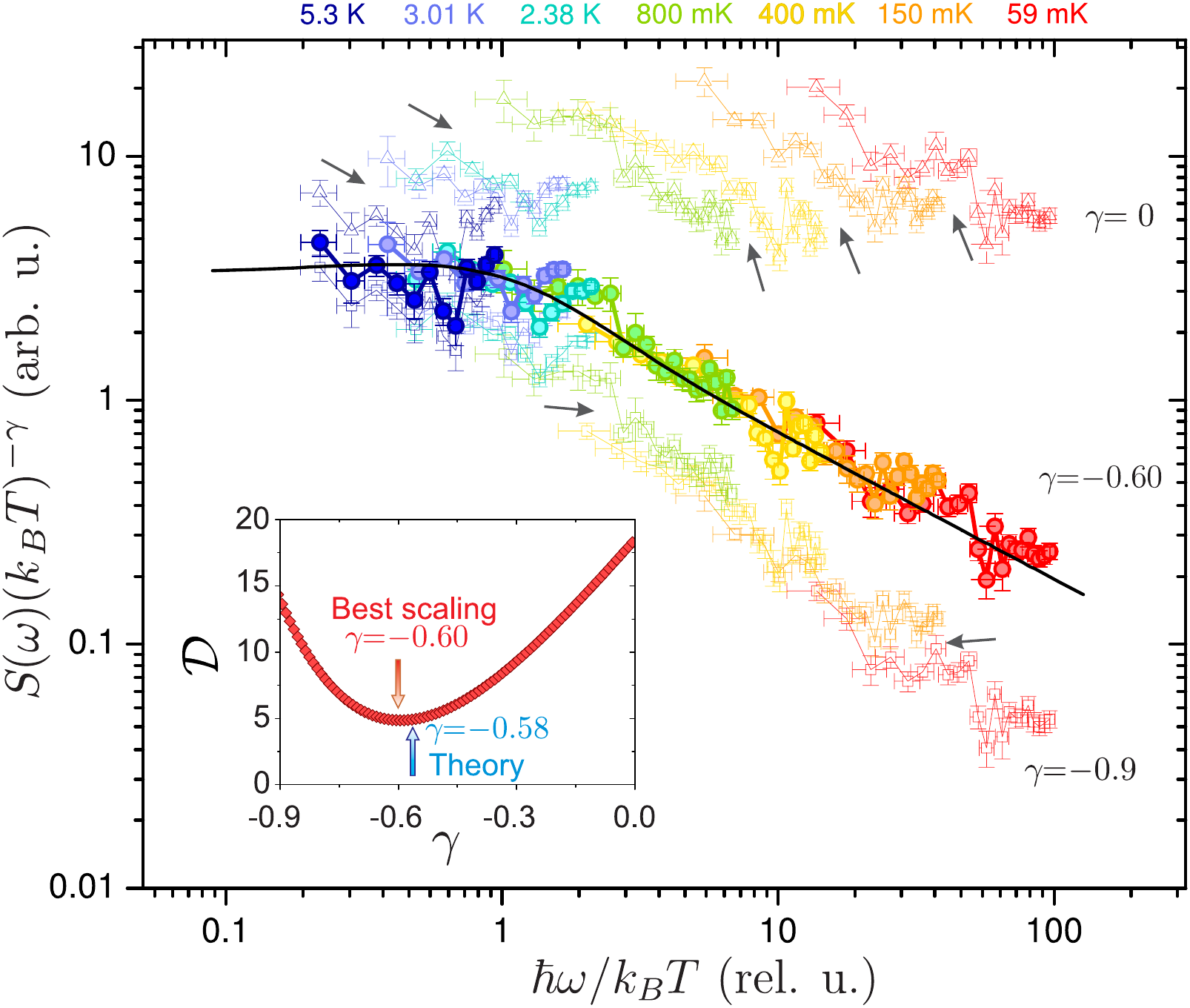}\\
		\caption{ From Ref.~\cite{Povarov2015}: Local dynamic structure factor $\mathcal{S}(\omega)$ measured in DIMPY at $H=9$~T at several temperatures, plotted in the scaling representation with different scaling exponents. Arrows mark the apparent
			violations of scaling for non-optimal exponent values. The solid line
			is the exact TLL scaling function  with the Luttinger parameter $K=1.25$.  Inset: measure of data overlap quality vs. scaling exponent. All data are from the LET spectrometer (Rutherford Appleton Laboratory, UK).\label{DIMPY2}}
\end{center} \end{figure}

As for magnetized spin chains, measuring the critical dynamics in the TLSL phase of DIMPY is very challenging due to a narrow measurement window. Consider the full excitation spectrum at $\mu_0H=7.5$~T shown in Fig.~\ref{DIMPY1} (from Ref.~\cite{Schmidiger2013-2,Schmidigerthesis}). Here the top and bottom panels show inelastic neutron scattering measurements and DMRG calculations, respectively. Most of the observed scattering has nothing to do with critical fluctuations. It either occurs at energies where the dispersion of excitations can no longer be approximated as linear (a prerequisite of TLSL behavior) or stems from gapped spectral components that were not involved at the soft-mode transition to the TLSL phase at $\mu_0H_c$. The only part of the spectrum where the true critical scattering can be observed unobstructed is below about 0.4~meV energy transfer. At the same time, one has to steer clear of the usual strong elastic incoherent scattering that is unavoidable in neutron experiments. Despite these difficulties, the measurement has been successfully carried out in a magnetic field $\mu_0H=9$~T using an extremely high resolution setup ($\delta E\sim 25~\mu$eV) \cite{Povarov2015,Schmidigerthesis}. Figure~\ref{DIMPY2} shows the resulting scaling plots for the dynamic structure factor. The three plots correspond to three different values of the temperature exponent and corresponds to $\frac{1}{2K}-1$ the present notation. For two of them there are clear ``tears'' in the data signifying poor data collapse. Optimal data collapse is obtained with $K=1.25(5)$ (central set of data points). The resulting scaling agrees very well with the theoretical scaling function (solid line) obtained from Eq.~\ref{TLSL:Sqw}. For the known exchange constants in DIMPY, DMRG predicts $K=1.2$ for $\mu_0H=9$~T  \cite{Povarov2015,Schmidigerthesis}.

\section{``Zero scale factor'' universality}
We now turn to QCPs in Heisenberg spin chains and ladders that are driven by an applied magnetic field. The most obvious example is the transition to the fully polarized state in high field in most non-ferromagnetic XXZ systems.  This case is also the easiest to describe, since the excitation spectrum at $H>H_\mathrm{sat}$ is gapped and {\em exactly} as given by spin wave theory. The magnon gap closes at the critical wave vector at precisely at $H=H_\mathrm{sat}$. There the dispersion is quadratic at low energies, so $z=2$ for this QCP.  Associating $\hat{S}^-$ with a magnon creation operator exactly maps the  problem into that of a hard core Bose gas. In this mapping the Boson density $\hat{\rho}$ corresponds to the reduction of magnetization  $NS-\hat{S}_{z,\mathrm{total}}$ and their chemical potential $\mu$ to $g\mu_B(H-H_\mathrm{sat})$. In three dimensions this famously allows to treat the saturation transition as a BEC of magnons \cite{Batyev1984}.  In $d=1$ below $H_\mathrm{sat}$ the system is a TLSL.

\begin{figure}
	\includegraphics[width=\columnwidth]{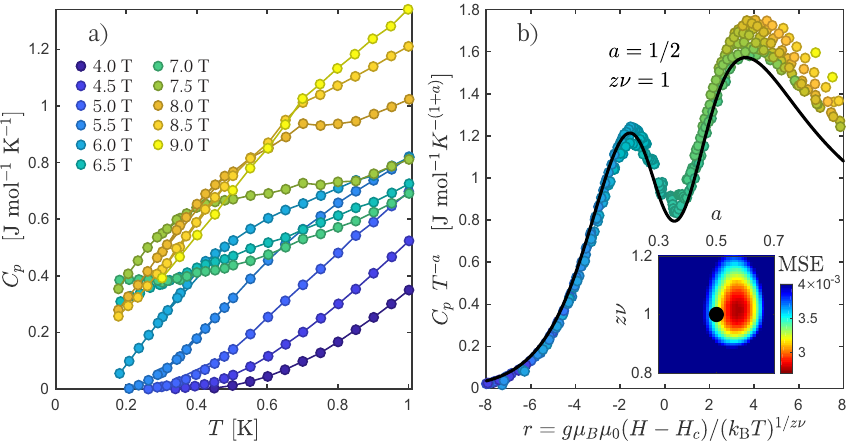}
	\caption{\label{BPCB1}From Ref.~\cite{Blosser2018}: a) Representative constant-field measurements of the magnetic contribution to specific heat in BPCB for a magnetic field along the $b$ axis. b) Symbols: measured magnetic specific heat for $0.17\leq T \leq 0.5$ plotted in scaled variables with the scaling exponents $a=1/2$ and $z\nu=1$. The solid line is the exact scaling function for free fermions in one dimension plotted with no adjustable parameters. Inset: False color plot of the empirical measure of the quality of the data collapse vs. the critical exponents. Optimal scaling is found for $a=0.57(10)$ and $b=1.01(10)$.}
\end{figure}

The consequences of the mapping to hard core bosons are even more profound in one dimension than for $d=3$ \cite{Sachdev1994}. This is because in  the low magnon density limit (in the vicinity of $H_\mathrm{sat}$) the properties of a $d=1$ hard core Bose gas are independent of interaction (e. g., radius of the hard spheres) and {\em completely universal}. The universality refers not only to the form of the scaling functions but even to the overall numerical coefficient \cite{Affleck1990-2,Affleck1991,Sorensen1993}. This is the so-called ``zero scale factor'' universality\cite{Sachdev1994}. All properties will be exactly as those for bosons with no interactions other than hard core repulsion, which in turn exactly map to {\em free} fermions via a Jordan-Wigner transform. Immediately this gives {\em exact} results for thermodynamic properties that are valid regardless of any details of the spin Hamiltonian \cite{Sachdev1994}. The exponents for Eqs.~\ref{scalingthermo} follow form $d=1$, $z=2$ and the fact that necessarily $z\nu=1$, since  the Zeeman term driving the transition commutes with the rest of the Heisenberg Hamiltonian \cite{Sachdev1994-2}. The exact scaling functions {\em including absolute prefactors} are as for free Fermions and trivial to compute:
\bea
\mathcal{M}(x)&=&g\mu_B\frac{Na}{V}\frac{\sqrt{2\pi m k_B }}{\pi \hbar}\times\nonumber\\
&\times& \int_0^\infty \dd y \frac{1}{\exp(y^2-x)+1},\\
\mathcal{C}(x)&=&k_B\frac{Na}{V}\frac{\sqrt{2\pi m k_B }}{\pi \hbar}\times\nonumber\\
&\times& \int_0^\infty \dd y \frac{(y^2-x)^2\exp(y^2-x)}{(\exp(y^2-x)+1)^{2}}.\label{cvz2}
\eea
Here $m$ is the magnon ``mass'', their dispersion at $H=H_c$ given by $\hbar\omega=\hbar^2q^2/2m$.

These universal results equally apply to field-induced gap closure transitions in dimerized AF spin chains, Haldane spin chains and AF spin ladders. In these cases the critical point once again separates a disordered gapped phase at $H<H_c$ from a gapless TLSL state at higher fields. The hard-core bosons correspond to the lowest-energy member of magnon triplet in the gapped phase, which goes soft at $H\rightarrow H_c$.  The predicted scaling for specific heat and susceptibility, as well as exact results from thermal expansion, magnetostriction and magnetocaloric effect have recently been beautifully put to the test in measurements on the $S=1/2$ chain compound CuPzN in the vicinity of $H_\mathrm{sat}$ \cite{Breunig2017}. Specific heat and magnetization scaling were also studied for the gap-closing transition in the spin ladder material \BPCB (BPCB) \cite{Blosser2018,Blosserthesis}. Fig.~\ref{BPCB1} shows the measured heat capacity in the vicinity of $H_{c}$ at different temperatures (a) and the corresponding scaling plot (b). The solid line is Eq.~\ref{cvz2} plotted without any adjustable parameters (the magnon mass is directly measured using inelastic neutron scattering). Any discrepancies are a result of the actual magnon dispersion deviating from a perfect parabola at higher energies \cite{Blosserthesis}.

\subsection{Dynamics}

\begin{figure}[h!]
	\includegraphics[width=\columnwidth]{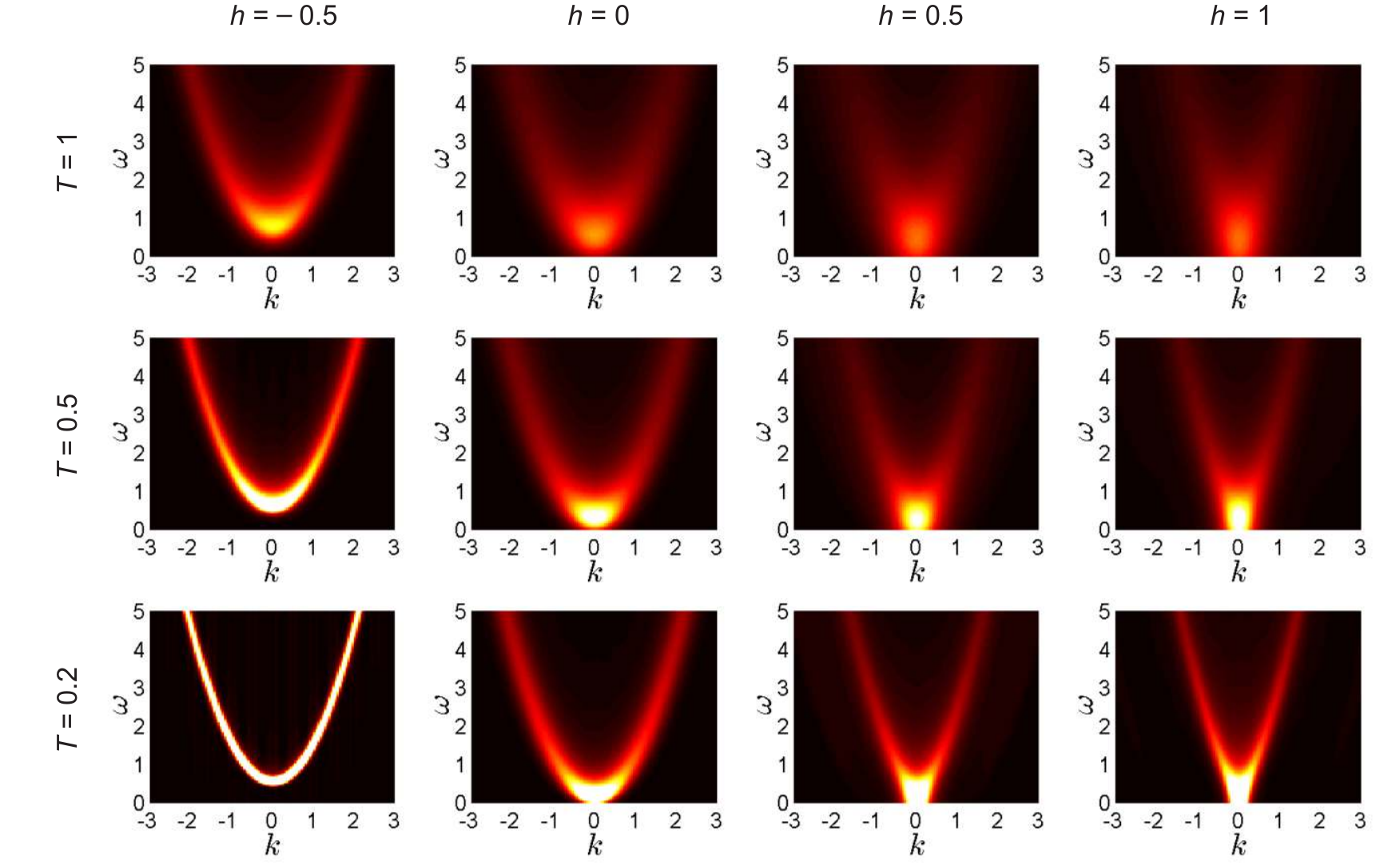}
	\caption{From Ref.~\cite{Blosser2017} \label{Fred} False color plots of the calculated transverse dynamic structure factor for impenetrable bosons in one dimensions at different chemical potentials $h=-0.5,0,0.5,1$, and temperatures $T=0.2,0.5,1$. }
\end{figure}

\begin{figure}[]
	\includegraphics[width=\columnwidth]{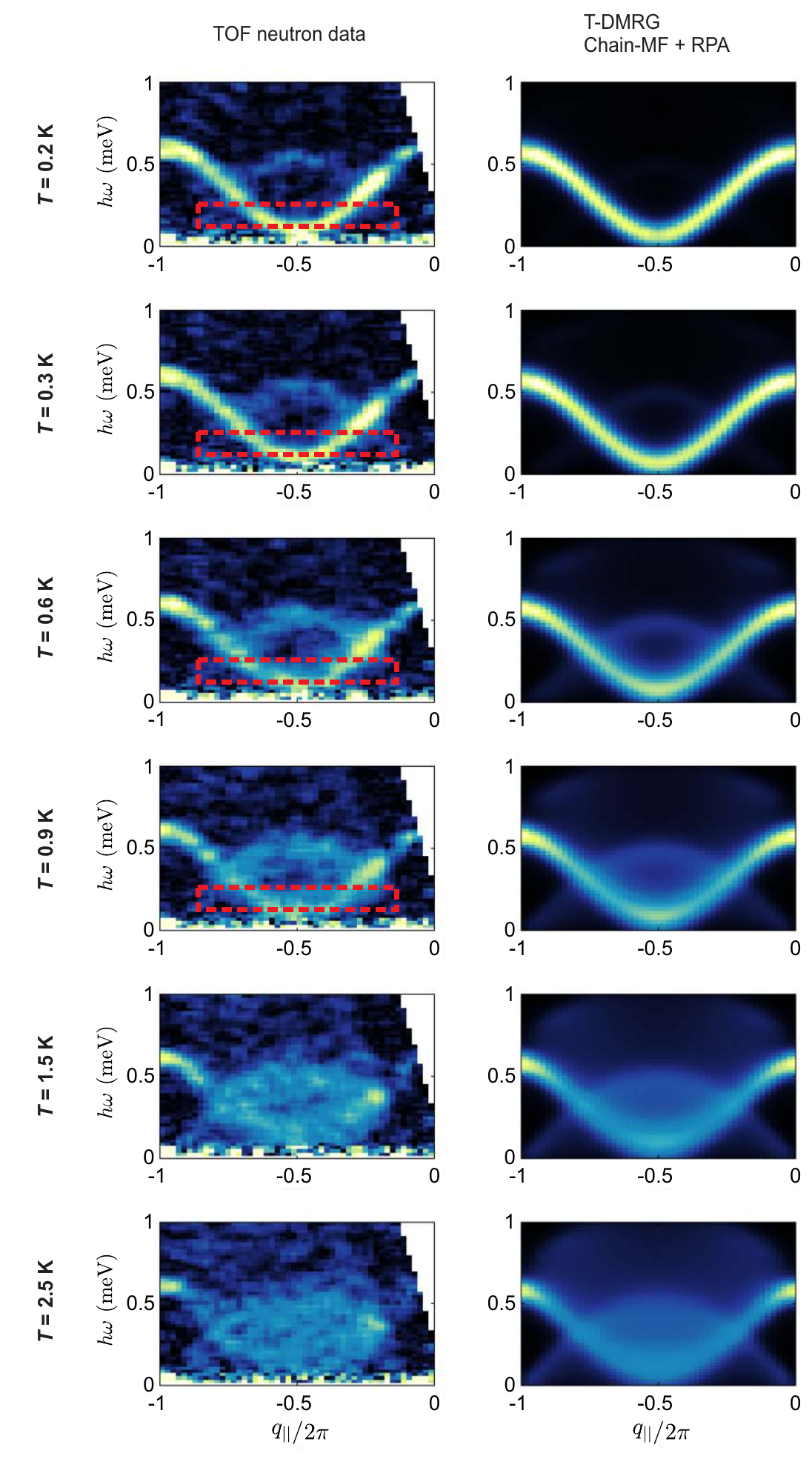}
	\caption{\label{KCu} From Ref~\cite{Blosser2017}: Measured and calculated spin excitation spectra of the Heisenberg spin chain compound K$_2$CuSO$_2$Cl$_2$ near saturation. The first column shows the inelastic neutron scattering data collected very close to saturation at $\mu_0 H = 4.5$~T  ($\mu_0 H_\mathrm{sat}=4.47$~T) at different temperatures. These plots correspond to slices integrated in the range $q_\perp/ 2\pi = 0 \pm 0.1$ and completely along the non dispersive direction $q_\mathrm{inter}$. The second column shows numerical results where inter-chain interactions are considered within a combined chain-MF and RPA approach. All data are from the time-of-flight instrument LET (Rutherford Appleton Laboratory, UK).}
\end{figure}

At this $d=1$ $z=2$ QCP the dynamic structure factor scales according to Eq.~\ref{SQWT} with $x=1/2$.  
Remarkably, the correlation function is known exactly\cite{Korepin1990}. Unfortunately, it has no simple form but is instead is expressed in terms of  Fredholm determinants. The asymptotic for $\tilde{\omega}\rightarrow 0$ and $\tilde{\omega}\rightarrow \infty$ are well known \cite{Sachdev1994}. In the general case it  needs to be computed numerically. The procedure is somewhat tedious but straightforward \cite{Blosser2017,Blosserthesis}. Typical results are shown in Fig.~\ref{Fred}.  The first experimental test of these predictions was attempted on the $S=1/2$ chhain material \KCUC which saturates in $\mu_0 H_\mathrm{sat}=4.47$~T \cite{Haelg2014,Haelgthesis}. Immediately it became clear that the critical fluctuations overlap with a highly structured continuum of non-critical scattering, which emerges as soon as one moves away from the QCP. This is illustrated in Fig.~\ref{KCu}. The left column shows neutron spectra measured at different temperatures at $H=H_\mathrm{sat}$.  Critical scattering is due to the magnons, whose cos-shaped dispersion  is clearly seen at the lowest temperature. The non-critical scattering appears as an ``inverted'' magnon dospersion at low temperatures and rapidly grows and forms a broad continuum at higher temperatures. Finite-temperature DMRG calculations revelaed that origin of these non-critical fluctuations: they are due to two-magnon states and are polarized parallel to the applied field. In contrast, critical fluctuations correspond to single-magnon excitations and therefore involve a spin flip. They are polarized transverse to the applied field.

\begin{figure}
	\includegraphics[width=\columnwidth]{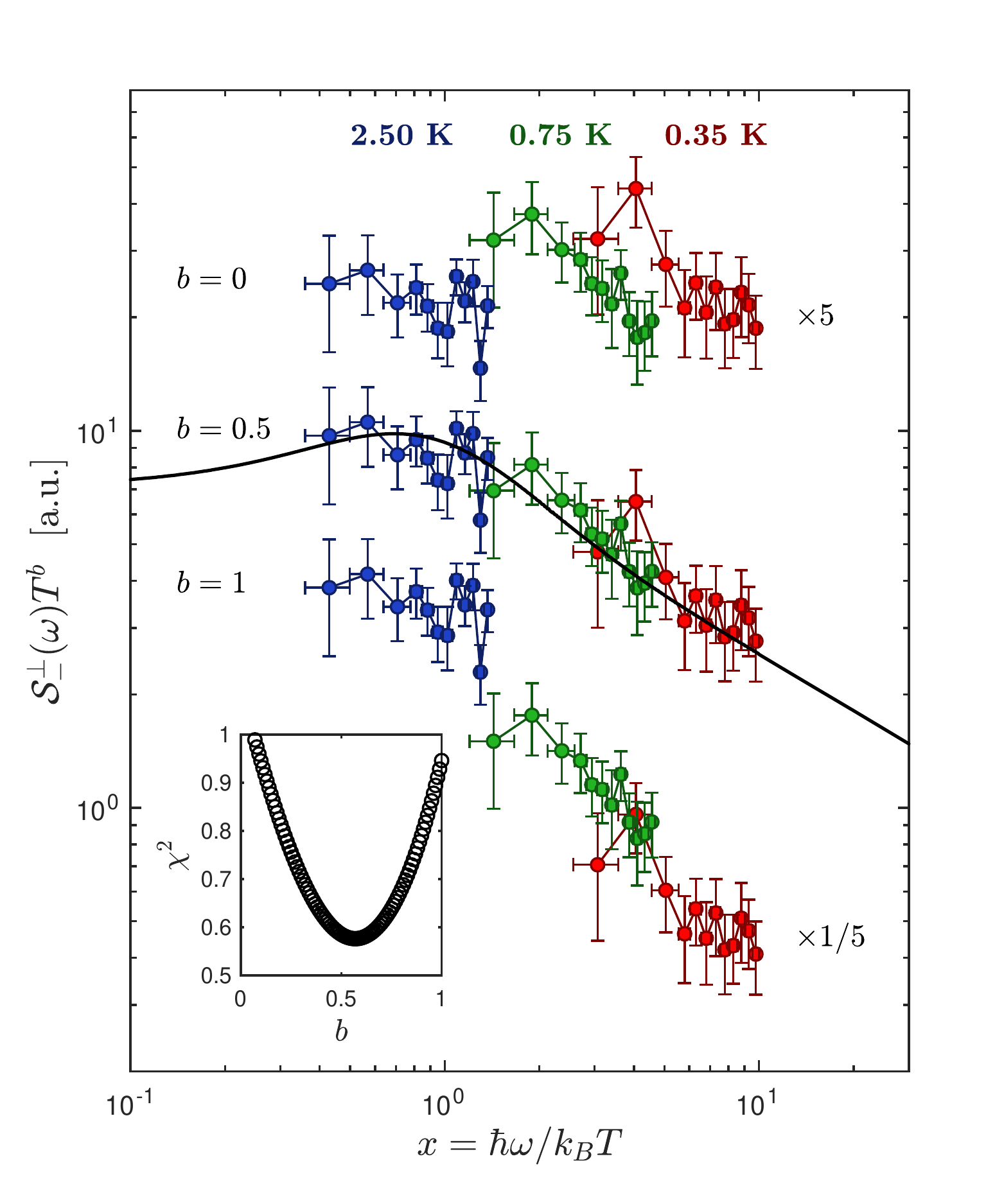}
	\caption{\label{BPCB2}From Ref.~\cite{Blosser2018}. Scaling plot of the antisymmetric transverse local dynamic structure factor  $\mathcal{S}_-^{\bot}(\omega)$ measured in BPCB   near the critical field at $H=6.5$~T. For the predicted scaling exponent $b=1/2$ all three data sets collapse onto a single continuous line. The inset shows the quality of the data collapse for different values of the critical exponent. Best scaling is found for $b=0.57(10)$. The solid line corresponds to the calculated scaling function. All data are from the time-of-flight instrument LET (Rutherford Appleton Laboratory, UK).}
\end{figure}

The challenge is to separate critical transverse scattering from the non-critical longitudinal one. In principle this could be done using polarized neutrons. As mentioned, this techniques usually involves a severe intensity penalty, making it impractical for measuring weak signals. There is, however a trick that can be employed in spin ladders. The Hamiltonian of such systems has an additional symmetry: the interchange of the two ladder legs. If the rungs are antiferromagnetic, the magnons are odd with respect to that symmetry\cite{Barnes1994,Bouillot2011}. This implies that single-magnon scattering is strongest for momentum transfers $q_\bot=\pi/l$ perpendicular to the leg direction, $l$  being the rung length in the material. In contrast, the two-magnon states will have $q_\bot=0$. The low-energy spectrum of a spin ladder near $H_c$ will be qualitatively similar to that of a spin chain neat $H_\mathrm{sat}$, with the difference that the critical and non-critical fluctuations will be separated in reciprocal space and can be measured independently. This idea was successfully implemented in measuring critical dynamics in the spin ladder material BPCB \cite{Blosser2018}. The resulting scaling plot for $\mathcal{S}(\tilde{\omega})$ is shown in Fig.~\ref{BPCB2}. The scaling exponent that optimizes the data overlap is $x=0.57(1)$ with the theoretical value $x=1/2$. The solid line shows the exact scaling function for free fermions.

\section{$d=1$ Ising model in transverse field}
The last example discussed here, the Ising model in a transverse field (IMFT), is perhaps the quintessential quantum phase transition\cite{Sachdevbook}. It occurs in all those situations described in the previous section (saturation, field-induced gap closure), but in the presence of anisotropy that breaks axial symmetry around the applied field direction. The transition is also of a soft-mode type: a magnon gap goes to zero at $H=H_c$. There are two key differences. First, the QCP separates two phases both of which are gapped \cite{Tsvelikbook}. Second, the Zeeman term doesn't commute with the rest of the Hamiltonian. The Zeeman energy is not a simple add-on to the magnon dispersion as in the axial case. Instead, the magnon dispersion as a ``relativistic'' form: $(\hbar\omega)^2=(\hbar u q)^2+g^2\mu_B^2(H-H_c)^2$ \cite{Sachdevbook,Tsvelikbook}. Unlike in the axial case, it is the ``mass'' of the magnon, not it's ``chemical potential'' that goes to zero at $H_c$. At precisely the critical field the dispersion is linear, ensuring $z=1$. At the critical wave vector the gap is linear with $(H-H_c)$ so we also have $z\nu=1$. At the QCP the system is similar to a TLSL in that it is gapless, has a linear spectrum and bears a fermionic description, specifically in terms of real (Majorana) fermions \cite{Tsvelikbook}. As a result,  correlation length goes as $1/T$ and the dynamic structure  factor is given by Eq.~\ref{TLSL:Sqw} with a particular value $K=4$ \cite{Sachdev1996,Sachdevbook}. Correspondingly, the temperature exponent in Eq.~\ref{SQWT} is $x=7/4$.

\begin{figure}[h!]
	\includegraphics[width=\columnwidth]{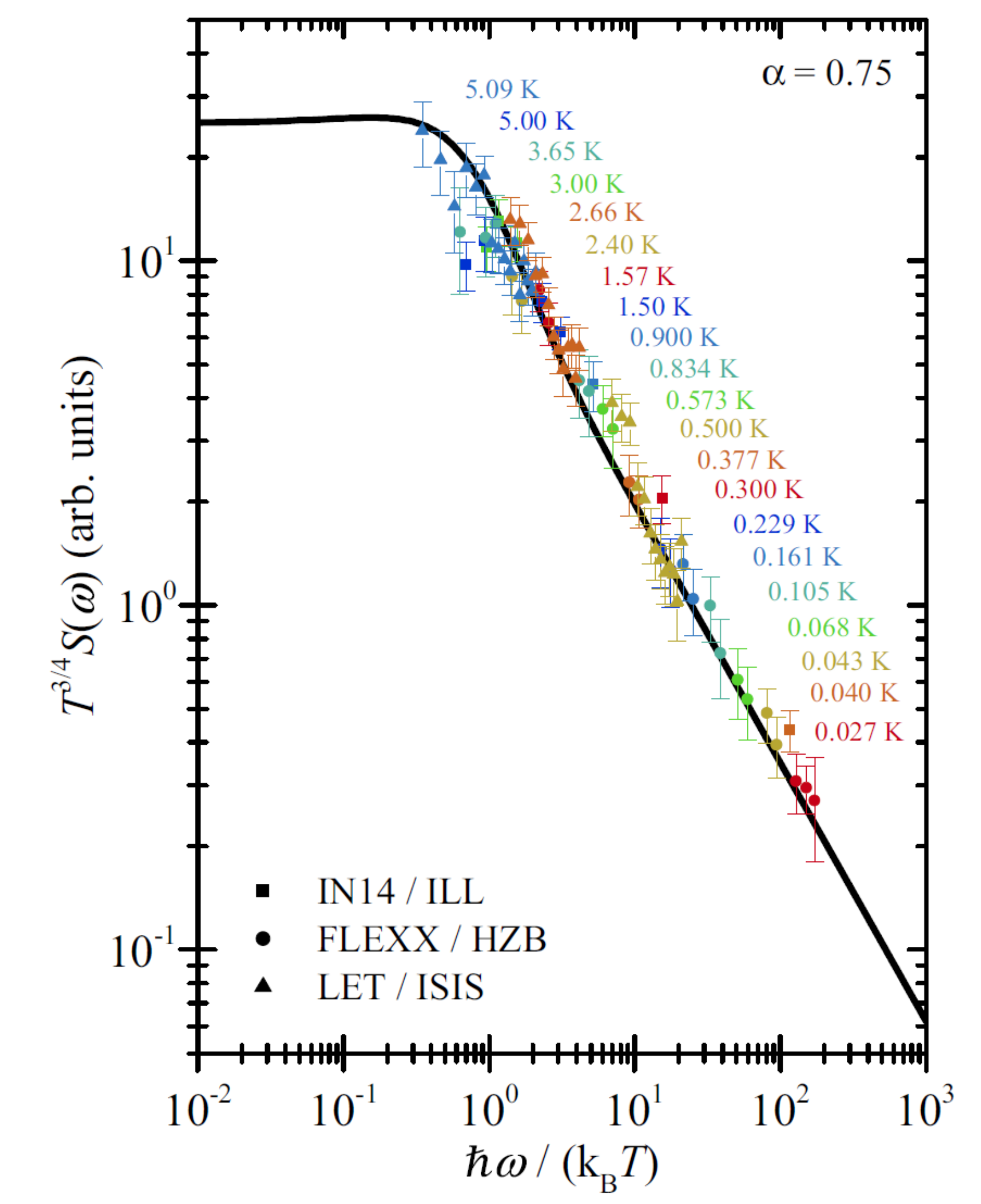}
	\caption{\label{NTENP}From Ref.~\cite{Haelg2015}. Scaling plot for the local dynamic
		structure factor measured in NTENP with the Ising scaling exponent
		$x-1 = 0.75$. The black line corresponds to the exact theoretical
		scaling function for the quantum critical point of an Ising chain in a
		transverse magnetic field. The experimental data are from 3-axis spectrometers IN-14 (Institut Laue-Langevin, Grenoble, France) and FLEXX (Helmholz Zentrum Berlin, Germany), and from the time-of-flight instrument LET (Rutherford Appleton Laboratory, UK). 
	}
\end{figure}

Although there are some stunningly beautiful experimental data on excitations in (quasi) one-dimensional Ising systems on both sides of the field-induced IMTF transition \cite{Coldea2010}, there were not many studies of finite-temperature critical dynamics. One such experiment was performed on the anisotropic  $S=1$ chain material \NTENP (aka NTENP). The exchange constants alternate between $J_1=2.1$~meV and $J_2=4.7$~meV, ensuring a dimerized singlet ground state, as opposed to a Haldane one \cite{Zheludev2004-2}. The system features strong single-ion anisotropy. As a result, the lowest-energy excitation triplet is split into a doublet with a gap $\Delta_1=1.07$~meV and a singlet at $\Delta_2=1.91$~meV. If a magnetic field is applied {\em perpendicular} to the anisotropy axis, the rotational symmetry is fully broken, resulting in an IMTF-type transition to what is at $T=0$ an AF ordered state at $\mu_0H_c\sim 11$~T \cite{Regnault2006}. The ``window'' where one can hope to measure the critical fluctuations is bounded above by the the central member of the magnon triplet at $\Delta_1$, which itself has nothing to do with critical scattering. Unfortunately, due to a previously unknown structural transition in this compound, there is a slight alternation of the $g$-tensor, which prevents the gap from fully closing: at $H_c$ it is still about $0.2$~meV \cite{Haelg2015,Haelgthesis}. This sets the lower bound for fluctuations that can be viewed as critical.

Despite these restriction, local dynamic structure factor measured in NTENP is in excellent agreement with theoretical predictions (Fig.~\ref{NTENP}). The corresponding scaling exponent $x-d$ was determined to be 0.77(2) compared to the expected value $3/4$. The only regret is that the data do not cover the all-important quantum relaxation regime $\tilde{\omega}\lesssim 1$. The fact that the data measured at the same frequency at different temperatures fall on a single line in the log-log plot of Fig.~\ref{NTENP} only means that scattering is temperature-independent at $k_BT\gg \hbar \omega$. However, the slope in the data collected at each temperature vs. energy is a non-trivial result: it establishes the power law behavior of $S(\omega)$ in the $T\rightarrow 0$ limit.

\section{Conclusion}

Tim Ziman (Institut Laue-Langevin) once jokingly commented that this type of activity is akin to ``measuring $\pi$'': you know exactly what you will get, as long you do the experiment right. This criticism is only partially justified. On the one hand, it is true that the experiments reviewed here test exact theoretical results that are long-standing and not really subject to any doubt. On the other hand, even $\pi$ should be measured a few times, just to make sure. It is such measurements performed in antique times that our unvavering faith in the applicability of geometry to the real universe is based upon. More important is another aspect. By its nature, scaling theory is expected to work well only very close to the QCP, at very small momenta and very small frequencies. The neutron studies reviewed here explore to what extent these results are relevant to real materials and to experiments with a realistic sensitivity and resolution.  We are not measuring $\pi$. We are checking if there are any {\em real} circles in Nature that can to some extent be approximated as being round.

\section{Acknowledgments}
Most of the new experimental material reviewed here was supported by the Swiss National Science Foundation, Division 2, and is the subject of successfully defended PhD dissertations at ETH Zurich, namely those of Dr. David Schmidiger \cite{Schmidigerthesis}, Dr. Manuel Haelg \cite{Haelgthesis}, Dr. Gediminas Simutis \cite{Simutisthesis} and Dr. Dominic Blosser \cite{Blosserthesis}. K. Povarov, S. Gvasaliya, W. Lorentz and D. Huevonen (ETH Z\"urich) also played an important role in many of the measurements. Which, in turn, would be impossible without the expert support of instrument scientists at neutron scattering user facilities: T. Perring, D. Voneshen, R. Bewley, H. C. Walker, D. T. Adjora, F. Demmel and T. Guidi (Rutherford Appleton Laboratory, UK);,  J. Robert and S. Petit (Laboratoire Leon Brillouin, CEA-CNRS, Saclay, France); M. Stone, A. I. Kolesnikov and A. T. Savichi (Oak Ridge National Laboratory, USA); L. P. Regnault and F. Bourdarot (CEA Grenoble, France); O. Sobolev (Forschungsneutronenquelle Heinz Maier-Leibnitz, Munich, Germany); N. P. Butch (National Institute of Standards and Technology, USA); D. L. Quintero-Castro (Helmholtz-Zentrum Berlin, Germany); A. Piovano and M. Boehm (Institut Laue-Langevin, Grenoble, France). While most samples for the described experiments were grown at ETH Z\"urich, the linear-chain cuprate crystals originate from the laboratories of Prof. B. Buechner (IFW Dresden, Germany), Prof. T. Masuda (The University of Tokyo, Japan) and Prof. A. Revcolevschi (Universite Paris-Sud, Orsay, France).

The present review is dedicated to the memory of Academician Andrei Stanislavovich Borovik-Romanov. The author is fortunate and privileged to have performed undergraduate and graduate studies under his guidance at the P. L. Kapitza Institute for Physical Problems in Moscow, Russia. Borovik-Romanov's seminal works on antiferromagnetism and spin waves have inspired the author's entire scientific career. The respect he indiscriminately showed to all people taught the author the meaning of being gentleman in science.

\end{document}